\documentclass[aps,prd,nofootinbib,twocolumn,superscriptaddress,preprintnumbers,balancelastpage,longbibliography]{revtex4-1}

\pdfoutput=1
\usepackage{subfigure}
\usepackage{graphicx}
\usepackage{epstopdf}
\usepackage{mathrsfs}
\usepackage{amssymb}
\usepackage{verbatim}
\usepackage{color}
\usepackage{multirow}
\usepackage{amsmath}
\usepackage{hyperref}
\usepackage{physics}
\usepackage{lipsum}
\usepackage{soul}

\usepackage{xcolor}

\hypersetup{pdfstartview=FitV,colorlinks=true,linkcolor=blue,citecolor=red,filecolor=black,urlcolor=blue}

\newcommand{\Trh}{T_{\rm RH}}
\newcommand{\Tfo}{T_{\rm FO}}

\newcommand{\beq}{\begin{equation}}
\newcommand{\eeq}{\end{equation}}
\newcommand{\bea}{\begin{eqnarray}}
\newcommand{\eea}{\end{eqnarray}}

\newcommand{\arh}{a_{\rm RH}}

\newcommand{\afo}{a_{\rm FO}}

\begin{document}

\preprint{UMN--TH--4513/25, FTPI--MINN--25/15}

\title{Ultra-Relativistic Freeze-Out: A Bridge from WIMPs to FIMPs}

\author{Stephen E. Henrich}
    \email[Correspondence email address: ]{henri455@umn.edu}
    \affiliation{William I. Fine Theoretical Physics Institute, School of Physics and Astronomy, University of Minnesota, Minneapolis, Minnesota 55455, USA}  

\author{Yann Mambrini}
     \email[Correspondence email address: ]{mambrini@ijclab.in2p3.fr}
\affiliation{Universit\'e Paris-Saclay, CNRS/IN2P3, IJCLab, 91405 Orsay, France}

\author{Keith A. Olive}
    \email[Correspondence email address: ]{olive@umn.edu}
    \affiliation{William I. Fine Theoretical Physics Institute, School of Physics and Astronomy, University of Minnesota, Minneapolis, Minnesota 55455, USA}


\begin{abstract} 
We re-examine the case for dark matter (DM) produced by ultra-relativistic freeze-out (UFO). UFO is the mechanism by which Standard Model (SM) neutrinos decouple from the  radiation bath in the early universe at a temperature $T_{d} \approx 1$ MeV. This corresponds to chemical freeze-out without Boltzmann suppression, such that the freeze-out (decoupling) temperature $T_{d}$ is much greater than $m_{\nu}$ and the neutrinos are therefore ultra-relativistic at freeze-out. While UFO has historically been rejected as a viable mechanism for DM production due to its association with hot DM and the accompanying incompatibility with $\Lambda$CDM,
we show that when the approximation 
of instantaneous reheating after inflation is lifted, UFO can produce cold DM and account for the entire observed relic density in large regions of parameter space. In fact, DM with masses ranging from sub-eV to PeV scales can undergo UFO and be cold before structure formation, given only a simple perturbative, post-inflationary reheating period prior to radiation domination. For some interactions, such as a contact interaction between the Higgs and DM scalars, there is a seamless transition between the WIMP and FIMP regimes which excludes UFO. However, for many other interactions, such as SM fermions producing fermionic DM via a heavy scalar or vector mediator, the WIMP to FIMP transition occurs \textit{necessarily} via a large intermediate region corresponding to UFO. We characterize the general features of UFO in this paper, while we supply a more detailed analysis in a companion paper. This mechanism is highly robust and does not require fine tuning. In particular, we find that UFO during reheating can produce the correct relic density ($\Omega_{\chi}h^2 = 0.12$) for DM masses spanning about 13 orders of magnitude, reheating temperatures spanning 17 orders of magnitude, and beyond the Standard Model (BSM) effective interaction scales spanning 11 orders of magnitude.  
\end{abstract}


\maketitle

\textit{Introduction--}Weakly interacting massive particles (WIMPs) and feebly interacting massive 
particles (FIMPs) are two paradigmatic candidates in the search for dark matter 
(DM) that are distinguished primarily by their mechanisms of production in the 
early universe, namely freeze-out and freeze-in respectively. While the WIMP 
scenario is well-motivated, it suffers from ever-tightening constraints from 
direct detection experiments \cite{LZ,PandaX:2024qfu,XENON:2025vwd,WIMPwaning}. On the 
other hand, direct and indirect detection efforts for FIMPs have encountered 
significant challenges due to extremely small couplings to Standard Model (SM) 
particles \cite{DawnFIMP}. In addition to the experimental challenges, there remain 
significant gaps in our theoretical understanding of the interface between 
these two mechanisms. Here, we describe a robust mechanism of cold DM production that 
lives in a large intermediate regime between WIMPs and FIMPs, namely 
ultra-relativistic freeze-out (UFO). Although UFO was first studied over 50 years ago \cite{GershteinZeldovich:1966, VysotskyDolgovZeldovich:1977}, it has often been neglected both historically and in recent work, in large part due to the belief that the resulting DM will always be hot 
or warm, and thereby disrupt structure formation \cite{HistoryofDM, ThermalDMBernal}.

Typically, WIMP freeze-out occurs well after the end of inflation after a phase of thermal equilibrium, erasing any memory of the physics of inflation and post-inflationary reheating. 
FIMP production may likewise be insensitive to the details of inflation and reheating, and in these instances it may be sufficient to assume that reheating takes 
place instantaneously when the scalar field driving inflation, the 
inflaton, decays. On the other hand, depending on the 
form of the interaction between DM and the SM, both WIMP and FIMP production 
may indeed be sensitive to the reheating process, in which case reheating 
cannot be assumed to be instantaneous \cite{Silva-Malpartida2,Giudice:2000ex,Garcia:2017tuj,gkmo1,gkmo2,Barman:2022tzk,HMO,ThermalDMBernal}. 
Here, we show that while UFO cannot produce cold relic DM under the assumption of instantaneous reheating, when this approximation is relaxed, the parameter space in which UFO can readily produce cold DM opens 
significantly. 

Some recent work has made progress toward characterizing the transition from 
WIMPs to FIMPs, using singlet scalar DM with a contact interaction to the SM 
as a case study \cite{Silva-Malpartida:2024emu,Silva-Malpartida2,Mondal:2025awq}. In these 
studies, the authors found that when the assumption of instantaneous reheating 
is relaxed, the values of the allowed WIMP couplings decrease while the 
allowed FIMP couplings increase, potentially rendering both paradigms more 
favorable from the standpoint of detection. 

We will show here that for a more general set of interactions, the WIMP to 
FIMP transition can be drastically different from the singlet scalar case. In particular, we will show that the transition between WIMPs and FIMPs exhibits 
two distinct but well-defined behaviors for interaction rates 
with either shallow or steep temperature-dependence. For example, consider a long-range interaction with a SM-DM cross section $\langle \sigma v \rangle \propto \frac{1}{T^{2}}$ and a rate $\Gamma \propto T$, for the former,
or for the latter an exchange of a massive mediator with $\langle \sigma v \rangle \propto \frac{T^{2}}{M^{4}}$ and $\Gamma \propto \frac{T^{5}}{M^{4}}$ at $M \gg T \gg m_{\chi}$, where $M$ is the mass of 
the heavy mediator and $m_{\chi}$ is the DM mass.
For rates with a 
steep temperature-dependence, there will always be an 
intermediate regime between WIMPs and FIMPs characterized by UFO, which is the 
central topic of our study. For other work on 
relativistic freeze-out see \cite{ArcadiLebedev:2019,SterileNeutrinoFOLebedev} and on relativistic freeze-in \cite{FreezeInUltra1,FreezeInRel1,FreezeInRel2}.

\textit{Classical UFO during radiation domination: neutrino decoupling--}Chemical freeze-out of DM in the early Universe occurs when the rate of 
number-changing interactions becomes too small to maintain equilibrium with the SM 
radiation bath. This happens when the interaction rate $\Gamma = \langle \sigma v \rangle n$ drops below the expansion rate characterized by the Hubble 
parameter, $H$. In the classical WIMP freeze-out scenario, the reason that $\Gamma$ drops below $H$ is because $\Gamma$ becomes exponentially suppressed 
by the Boltzmann factor ($e^{-m_{\chi}/T}$) when the temperature drops below $T \approx m_{\chi}$. However, this is not the only way for particles to 
freeze-out. Instead, it is possible for $\Gamma$ to drop below $H$ without any 
exponential suppression. In this case, $\Gamma(T)$ can simply drop below $H(T)$ due to its temperature dependence being steeper than the temperature 
dependence of $H(T)$. Indeed, this is precisely what happens in the case of 
(left-handed) neutrino decoupling. Neutrinos decouple when the radiation 
temperature is $\mathcal{O} $(MeV); and because the decoupling temperature is far 
above the neutrino masses, there is no Boltzmann suppression. Yet, freeze-out 
still occurs because $\Gamma(T)$ is more steeply $T$-dependent than $H(T)$. In particular, since $\Gamma = \langle \sigma v \rangle n_{\rm eq}$ with $n_{\rm eq} \propto T^{3}$ and $\langle \sigma v \rangle \propto \frac{T^{2}}{M^{4}_{W}}$, we have $\Gamma \propto T^{5}$. In contrast, $H \propto T^{2}$ during radiation domination. This 
relative temperature dependence between $\Gamma$ and $H$ thus leads to 
neutrino decoupling (freeze-out) without Boltzmann suppression, and the 
decoupled neutrinos are ultra-relativistic at freeze-out since the freeze-out 
temperature $T_{\rm FO} \gg m_{\nu}$.

The late relativistic freeze-out of neutrinos at $T\sim 1$~MeV, leads to very strong constraints on the sum of light neutrino masses \cite{Gershtein:1966gg,Cowsik:1972gh,Szalay:1974jta}, $\Sigma m_\nu \le 11$~eV for $\Omega_\nu h^2 \le 0.12$,
where $\Omega_\nu$ is the fraction of critical density to close the Universe and $h = 100~H_0$~km/Mpc/s is the scaled present day Hubble parameter. If saturated, neutrinos would account for the DM in the Universe \cite{SS}. A more quantitative discussion of UFO in this context can be found in \cite{HGMO}. However, this would be hot DM
and the constraints from the shape of the matter power spectrum combined with CMB data are considerably stronger $\Sigma m_\nu \lesssim 0.1$~eV \cite{Planck:2018vyg,Brieden:2022lsd} 
precluding SM neutrinos from making up the dark matter. 

\textit{Conditions required for UFO--}We can extrapolate from the particular case of neutrino decoupling to observe a general principle. For $\Gamma(T) \propto T^{\gamma_{1}}$ and $H(T) \propto T^{\gamma_{2}}$ where $\gamma_{1} > \gamma_{2}$ and $\Gamma>H$ at early times, $\Gamma$ will \textit{always} eventually drop below $H$ at low enough temperatures, leading to freeze-out without Boltzmann suppression, for sufficiently low $m_\chi$. Such a process 
will always correspond to UFO. The condition on the temperature-dependence of $\Gamma$ required for UFO to be possible is $\gamma_{1} > \gamma_{2}$, or equivalently \begin{equation}
\frac{d \ln \Gamma_{\rm rel}}{d \ln T} > \frac{d \ln H}{d \ln T}\,,
\label{Eq:UFOcondition}
\end{equation}
\noindent where $\Gamma_{\rm rel}$ is the interaction rate in the relativistic regime. So, for instance, neutrino decoupling is possible since $\frac{d \ln \Gamma_{\rm rel}}{d \ln T} = 5$ for the weak interactions below the electroweak scale and $\frac{d \ln H}{d \ln T}=2$ during radiation domination.

This condition can be applied to DM freeze-out rather than neutrino decoupling. Consider a generic BSM interaction characterized by a thermally averaged cross section of the form 
\beq
\langle \sigma v \rangle = \frac{T^{n}}{\Lambda^{n+2}}\,,
\label{crossSection}
\eeq 
for $T \ll \Lambda$. This parametrization is applicable to a broad range of cross sections in the (ultra)relativistic regime, where $T \gg m_{\chi}$. With this parametrization, our DM production rate becomes \beq
\Gamma_{\rm rel}=n^{\rm eq}_{\chi}\langle \sigma v \rangle\,,
\label{gammaRel}
\eeq 
\noindent where the equilibrium number density of DM in the (ultra) relativistic regime is given by \beq
n^{\rm eq}_{\chi}=\frac{g_{\chi} \zeta(3)}{\pi^2}T^{3}\,,
\label{numDens}
\eeq
where $g_{\chi}$ is the number of degrees of freedom of the DM particle $\chi$. Because we are interested in scenarios in which freeze-out is driven by the relative temperature dependence of $\Gamma$ and $H$ rather than Boltzmann suppression, we require $m_{\chi} < T_{\rm FO}$,  where $T_{\rm FO}$ is defined by $\Gamma(T_{\rm FO}) \simeq H(T_{\rm FO})$, so that DM will be relativistic at freeze-out.

To obtain a useful form of Eq.~(\ref{Eq:UFOcondition}), we require the Hubble 
parameter as a function of temperature for each cosmological epoch under 
consideration.  Radiation domination is characterized by an equation of state $p/\rho = 1/3$, where $p$ and $\rho$ are the pressure and energy density of 
the radiation.  During radiation domination,  $H(T)$ is given simply by
\beq
H_{\rm rad}(T)=\sqrt{\frac{\alpha}{3}}\frac{T^{2}}{M_{P}}\,,
\label{HubbleRad}
\eeq
\noindent where $\alpha=\frac{\pi^2}{30}g^{*}$, $g^{*}$ is the number of 
relativistic degrees of freedom in the SM radiation bath and $M_{P}\simeq2.4 \times 10^{18}$~GeV is the reduced Planck mass. However, during the reheating era, the 
Hubble parameter does not take this form since the energy budget of the 
universe is dominated by the inflaton's energy density rather than radiation. We consider here perhaps the simplest and most well-studied reheating scenario 
in which the inflaton transfers its energy to SM radiation via harmonic 
oscillations about the minimum of a quadratic potential, $V(\phi)\propto \phi^2$. In this case, the equation of state is simply $p/\rho=0$, where $p$ 
and $\rho$ are now the pressure and energy density of the oscillating inflaton 
condensate. Importantly, we consider reheating in the context of inflation models which are constrained to be consistent with CMB observables, such as the amplitude of the primordial power spectrum ($A_{s}\approx 2.1 \times 10^{-9}$), the scalar spectral index ($n_{s}\approx 0.965$), and the tensor-to-scalar ratio ($r\lesssim0.036$) \cite{Planck:2018vyg}. In particular, we use T-attractor models consistent with Starobinsky-type inflation, which is favored by CMB measurements \cite{Planck2018:Inflation}. For additional details, see our companion paper \cite{HGMO} along with \cite{gkmo1,gkmo2,HMO}. For reheating, we consider inflaton decay to either fermions or bosons. In this case, the Hubble parameter 
can be expressed in terms of the temperature of the 
newly forming radiation bath during the reheating era \cite{kmo}:
\beq
H_{\rm RH}(T)=\sqrt{\frac{\alpha}{3}}\frac{T^{4}}{T_{\rm RH}^{2} M_{P}}\,.
\label{HubbleEMD}
\eeq
Combining Eqs.~(\ref{crossSection}), (\ref{gammaRel}) and (\ref{numDens}), 
the condition required for UFO given by Eq.~(\ref{Eq:UFOcondition}) for a generic thermally averaged BSM cross section and a general cosmological era with $H(T) \propto T^{\gamma_{2}}$, we find that UFO is only possible for cross sections satisfying
\beq
n > \gamma_{2}-3\,.
\label{UFOnCondition}
\eeq

\noindent This implies that cross sections with $n>-1$ are required for UFO to 
be possible during a radiation dominated era, while for a reheating 
period with $\gamma_{2}=4$ we require $n>1$. Eq.~(\ref{UFOnCondition}) 
concisely summarizes our earlier assertion that UFO is only possible for cross 
sections with sufficiently steep temperature dependence. An immediate 
consequence of this is that cross sections of the form $\langle \sigma v \rangle \propto \frac{1}{T^{2}}$ such as a scalar contact interaction (as in the 
singlet scalar case) are not consistent with UFO for any typical cosmological 
era. 
In contrast, most interactions involving heavy mediators with $m_{\chi}, T \ll M$ are amenable to UFO. We further discuss and illustrate the distinction between these two types of interactions in the End Matter Section \ref{EMA}.

\textit{Exploring the freeze-out/freeze-in transition--} The UFO regime naturally and inevitably emerges in the transition between WIMPs and FIMPs for cross sections satisfying Eq.~(\ref{UFOnCondition}). Consider an interaction for which DM comes into equilibrium with the SM radiation bath at early times and later freezes out due to Boltzmann suppression (i.e. WIMP-like freeze-out). We can then imagine gradually reducing the coupling(s) while keeping the DM mass fixed, until we find that the DM no longer reaches equilibrium at any point in the universe's history. In this latter case, we will be in the FIMP regime and the relic abundance will be determined by freeze-in. At couplings slightly greater than the critical coupling, we will discover two distinct possibilities, depending on the interaction. One possibility corresponds to WIMP-like non-relativistic freeze-out, and one to UFO.

\begin{figure*}[ht!]
\centering
\vskip .2in
\includegraphics[width=\textwidth]{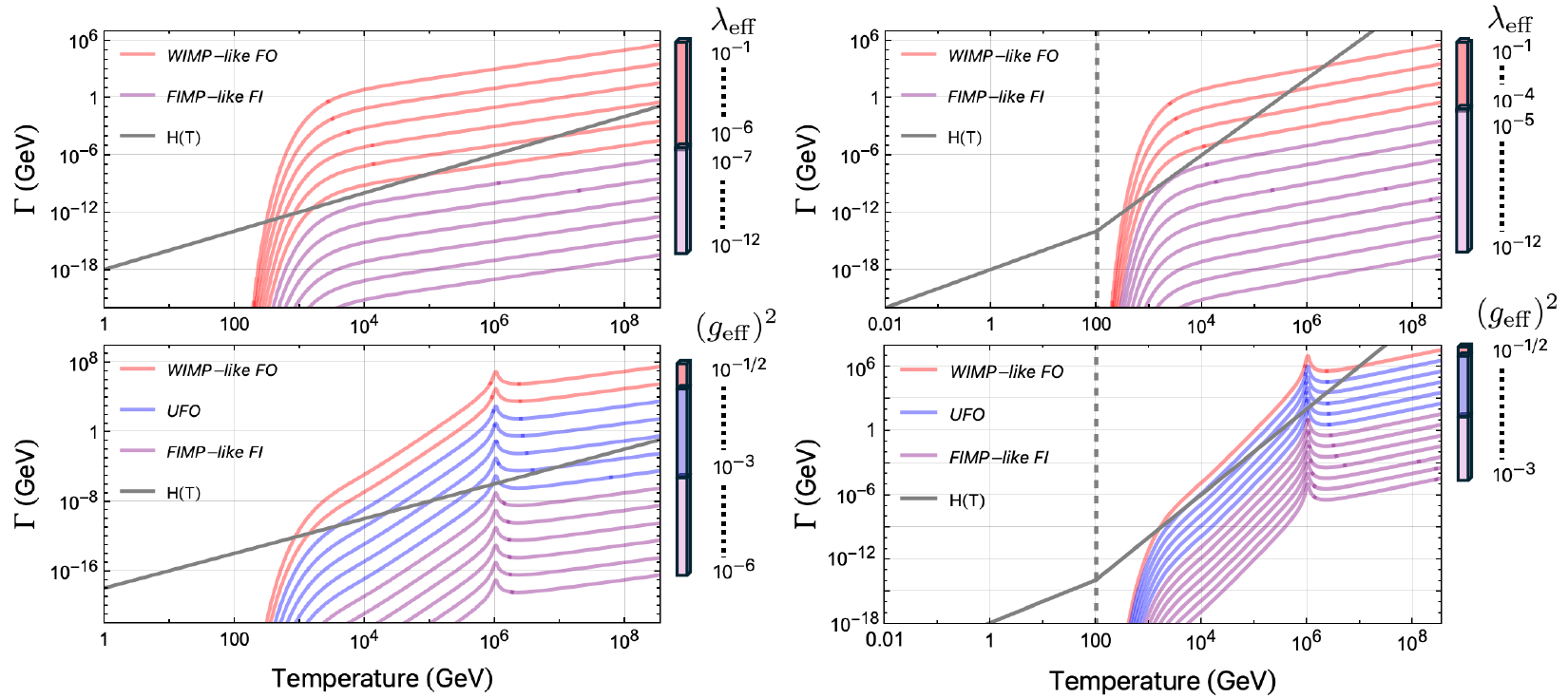}
\caption{\em \small $\Gamma$ vs. $T$ plots illustrating the transition from WIMP-like to FIMP-like behavior for two different interactions and high temperature instantaneous reheating (left) vs. non-instantaneous reheating with $\Trh = 100$~GeV (right). Freeze-out occurs at the low-temperature intersection of $\Gamma$ and $H$ (for both WIMP-like and UFO DM). UFO is observed for the FSF (or FVF) interaction (bottom panels) but not for the contact interaction between scalar DM and the Higgs (top panels). The vertical gray dashed lines correspond to $T=T_{\rm RH}$. Results are shown for $m_{\chi}=10^4$~GeV and $M_{s}=10^{6}$~GeV.}
\label{WIMPtoFIMP}
\end{figure*}

\begin{figure}[!ht]
\centering
\vskip .2in
\includegraphics[width=3.3in]{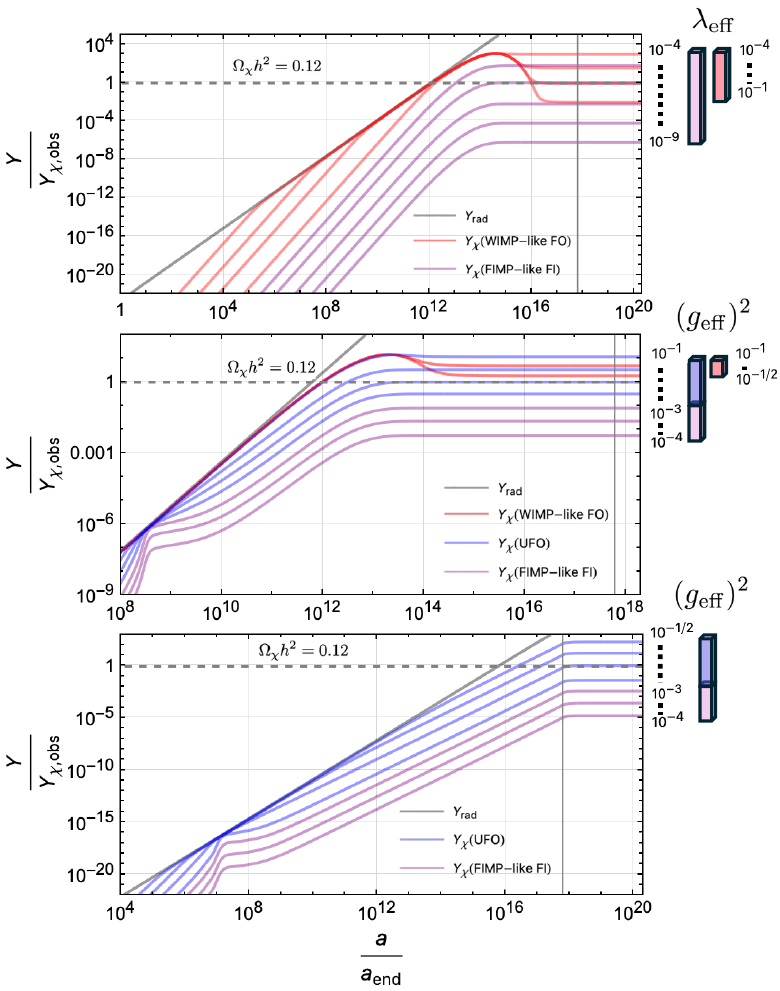}
\caption{\em \small Evolution of the comoving DM number density ($Y_{\chi}$) during reheating, illustrating the FO/FI transition for three different cases. \textbf{Top panel:} A contact interaction between DM and SM scalars for $m_{\chi}>\Trh$. UFO is absent. \textbf{Middle panel:} FSF/FVF interaction for $m_{\chi}>\Trh$. WIMP-like FO, UFO, and FI are all observed. \textbf{Lower panel:} FSF/FVF interaction for $m_{\chi}<\Trh$. Only UFO and FI are observed. 
The vertical gray lines correspond to $\frac{a_{\rm RH}}{a_{\rm end}}$, with $T_{\rm RH}=100$~GeV.}
\label{Ychi}
\end{figure}

We illustrate the WIMP to FIMP transition in precisely this manner in Figs.~\ref{WIMPtoFIMP} and \ref{Ychi}. We consider two BSM interactions between DM and SM particles, which are representative of the distinctive behavior for rates with mild vs. steep temperature-dependence. For the rate with mild temperature dependence, we consider a contact interaction between DM scalars and the Higgs. The thermally averaged cross section for this interaction takes the form $\langle \sigma v \rangle = \beta_{1} \frac{\lambda^{2}}{T^{2}}=\frac{\lambda_{\rm eff}^{2}}{T^2}$ for $T \gg m_{\chi}$ such that $n=-2$, where $\lambda$ is the relevant coupling, and we have defined $\lambda_{\rm eff}=\sqrt{\beta_{1}}\lambda$. For the rate with a steep temperature-dependence, we consider fermionic DM produced by SM fermions via a heavy scalar or vector mediator (FSF or FVF). The thermally averaged cross section for this interaction takes the form $\langle \sigma v \rangle = \beta_{2} g_{1}^2g_{2}^2 \frac{T^{2}}{M^{4}}$ for $M \gg T \gg m_{\chi}$ such that $n=2$. The couplings for the FSF/FVF interaction are $g_{1}$ and $g_{2}$, and we define $g=(g_{1}^{2}g_{2}^{2})^{1/4}$ and $g_{\rm eff}=(\beta_{2})^{\frac{1}{4}} g$, such that $\langle \sigma v \rangle=g_{\rm eff}^{4}\frac{T^2}{M^4}$. The constants $\beta_{i}$ are numerical factors from thermal averaging. 

In Fig.~\ref{WIMPtoFIMP}, we plot the DM production rate $\Gamma(T)$ for each interaction in the case of a standard radiation-dominated era assuming instantaneous reheating (left panels) and for non-instantaneous reheating (right panels). 
The gray diagonal lines correspond to the Hubble rate with $\gamma_2 = 2$ (left panels) and $\gamma_2 = 2, 4$ (right panels). 
In each case, a broad range of choices of the effective coupling are depicted while $m_{\chi}$ is fixed at $10^{4}$~GeV.

We also illustrate the FO/FI transition for these two interactions in Fig. \ref{Ychi}, this time depicting the evolution of the comoving DM number density, $Y_{\chi} = n_\chi a^3$, where $a$ is the cosmological scale factor. Equilibrium of the DM with the SM radiation bath is maintained so long as $\Gamma > H$, i.e. for portions of the curves in Fig.~\ref{WIMPtoFIMP} which are above the diagonal line. This corresponds to the portions of curves in Fig.~\ref{Ychi} which are merged with the gray diagonal line, which represents the equilibrium comoving number density, $Y_{\rm eq}$.

For the contact interaction at strong coupling and instantaneous, high temperature reheating, we see that WIMP-like freeze-out occurs as expected due to the exponential suppression of the equilibrium number density of DM particles (red curves, top left panel of Fig.~\ref{WIMPtoFIMP}). In contrast, for smaller couplings, the DM will never reach equilibrium and we are in the FIMP regime (purple curves, top left panel of Fig.~\ref{WIMPtoFIMP}). When we relax the assumption of instantaneous reheating, these same properties are still observed (Fig.~\ref{WIMPtoFIMP}, top right panel and Fig.~\ref{Ychi}, top panel), albeit for slightly different couplings. Note that there is no intermediate regime corresponding to UFO for the contact interaction either in the radiation dominated era or during reheating, as illustrated for instance in the top panel of Fig.~\ref{Ychi} by the lack of blue curves separating from the diagonal line ($Y_{\rm eq}$). This is consistent with our conditions described above in Eq.~(\ref{UFOnCondition}), since $n=-2$ for the contact interaction.

The situation is strikingly different for the FSF (or FVF) interaction. For FSF/FVF at very low couplings for either instantaneous or non-instantaneous reheating, we are in the FIMP regime (purple curves, bottom panels of Fig.~\ref{WIMPtoFIMP}). However, if we increase the coupling such that the interaction reaches equilibrium (seen by the evolution curves which cross (merge with) the diagonal line in Fig.~\ref{WIMPtoFIMP} (Fig.~\ref{Ychi})), there will \textbf{\textit{not}} be a transition to the WIMP regime with non-relativistic or semi-relativistic freeze-out. Instead, a steady increase in the coupling for FSF/FVF (and any other interaction satisfying Eq.~(\ref{UFOnCondition})) leads to a transition from the FIMP regime to \textit{ultrarelativistic} freeze-out (blue curves, bottom panels of Figs.~\ref{WIMPtoFIMP} and \ref{Ychi}). It is worth noting that if we steadily increase the coupling for FSF/FVF from the FIMP regime to just above the critical coupling, we first encounter a regime where equilibrium is achieved only briefly near $T=M$ due to the resonance peak. At even stronger couplings, we find a regime where UFO will occur with $\Gamma \propto T^{5}$. In this region, it is easy to show using Eqs.(\ref{crossSection}), (\ref{gammaRel}), (\ref{numDens}) and (\ref{HubbleEMD}) that the ultrarelativistic freeze-out temperature is given by
\begin{equation}
T_{\rm FO}=\sqrt{\frac{\alpha}{3}}\frac{3 \pi^2}{2g_{\chi} \zeta(3)} \frac{\Lambda^4}{\Trh^2 M_{P}}.
\label{TUFO}
\end{equation}
Indeed, one can see from the bottom panels of Fig.~\ref{WIMPtoFIMP}, and the middle panel of Fig.~\ref{Ychi} that there is a large range of couplings corresponding to UFO before we eventually reach the WIMP regime at even stronger couplings, where freeze-out will be driven by Boltzmann suppression. However, for some choices of $m_{\chi}$, $T_{\rm RH}$, and $\Lambda$, only UFO and FI are possible, but not WIMP-like freeze-out. This is the case for the bottom panel of Fig.~\ref{Ychi}, where $m_{\chi}=1$~MeV. In this case, we encounter the unitarity bound \cite{Bernal:2023ura} before WIMP-like freeze-out can be attained (thus, there are no red curves in Fig.~\ref{Ychi}, bottom panel). While Figs.~\ref{WIMPtoFIMP} and ~\ref{Ychi} illustrate the WIMP-to-FIMP transition, they do not convey the breadth of UFO parameter space consistent with $\Omega_{\chi}h^2=0.12$. We derive and report the allowed UFO parameter space in the End Matter Section \ref{EMParamSpace}.

Finally, we address the important requirement that UFO DM must not be hot, and must not be too warm by the time of structure formation. The warmness constraint derived from the Lyman-$\alpha$ forest \cite{Irsic:2017ixq} 
requires that the typical velocity of DM
at the time of structure formation should be highly non-relativistic, $v_\chi < 2 \times 10^{-4}$ at $T \simeq 1$~eV. 
Taking $v_\chi = p_\chi/m_\chi$ and $p_\chi \simeq \Tfo(\afo/a)$, we can redshift the DM
momentum after UFO through both the reheating and radiation-dominated epochs using 
$p_\chi \simeq \Tfo  \left( \frac{\afo}{\arh} \right) \left({\frac{\arh}{a}}\right)$
so that $p_\chi \simeq \Tfo  \left( \frac{\Trh}{\Tfo} \right)^{\frac83} \left( \frac{T}{\Trh} \right)$
and at 1 eV, we have the constraint that $m_\chi > 5~{\rm keV} \left( \frac{\Trh}{\Tfo} \right)^{\frac53}$.

Since we consider UFO during reheating, such that $T_{\rm FO}> \Trh $, we find that UFO DM with mass $m_{\chi}>5 \text{ keV}$ is automatically cold by structure formation. This lies in stark opposition to the conventional supposition that neutrino-like FO necessarily produces hot or warm DM. In fact, DM with much smaller mass (e.g. 1 eV) can become cold under this constraint if $\Tfo \gg \Trh$. The constraint above requires some care regarding whether the out-of-equilibrium production after UFO is UV vs. IR dominated, which we address in our companion paper \cite{HGMO}. Other important constraints are the effective number of relativistic degrees of freedom, $N_{\rm eff}$, as well as the lower bound on $T_{\rm RH}$ from BBN. \cite{ysof} UFO DM easily satisfies the current limits of $\Delta N_{\rm eff} \lesssim 0.18$ (see \cite{HGMO} for more details); and we have applied a cut-off to our parameter space respecting the BBN bound of $T_{\rm RH}>$~4 MeV. We note that for baryogenesis, reheating temperatures above $\mathcal{O}(100)$~GeV are preferable.

\textit{Conclusion--} We have demonstrated in a straightforward manner that there are two fundamentally distinct ways that the WIMP and FIMP regimes can be connected. Either 1) WIMP-like freeze-out can transition directly to freeze-in without an intermediate UFO regime; or 2) the transition from WIMPs to FIMPs can occur via a large intermediate, and previously under-investigated, regime corresponding to UFO. It is the temperature-dependence of the DM-SM interaction rate along with the temperature-dependence of the Hubble parameter that distinguishes these two possibilities, according to Eq.~(\ref{UFOnCondition}). Moreover, it is critical to consider UFO because there are some circumstances in which WIMP-like FO is prohibited due to unitarity, but UFO and FI are both permitted. While we have investigated UFO during reheating, this mechanism is also compatible with a non-standard cosmological era of Early Matter Domination (EMD).  \cite{Silva-Malpartida:2024emu,Barman:2024,Banerjee:2022} However, cold UFO DM does not require entropy injection beyond that which is necessary to reheat the universe after inflation.

While UFO DM is highly constrained if one makes the assumption of instantaneous reheating \cite{HGMO}, relaxing this assumption immediately opens up large regions of parameter space in which UFO will be the operative process that determines the relic abundance of cold DM for many well-motivated BSM processes such as vector portal interactions (e.g. $Z'$ portals and dark photons), moduli portals, massive spin-2 portals, GUT models such as those based on SO(10)~\cite{SO10} , high-scale and low-scale SUSY models, spin-3/2 decay and all non-renormalizable effective interactions. While we've elected to use a simple model of perturbative reheating for the present study, our discovery that UFO can produce cold DM is also applicable to many other reheating scenarios.

\section*{Acknowledgements} \label{sec:acknowledgements}
  This project has received support from the European Union's Horizon 2020 research and innovation program under the Marie Sklodowska-Curie grant agreement No 860881-HIDDeN.
  The work of K.A.O.~was supported in part by DOE grant DE-SC0011842  at the University of
Minnesota. Y.M. acknowledges support by Institut Pascal at Université Paris-Saclay during the Paris-Saclay Astroparticle Symposium 2024, with the support of the P2IO Laboratory of Excellence (program “Investissements d’avenir” ANR-11-IDEX-0003-01 Paris-Saclay and ANR- 10-LABX-0038), the P2I axis of the Graduate School of Physics of Université Paris-Saclay, as well as the CNRS IRP UCMN.

\clearpage

\section*{End Matter}

\subsection{Schematic Comparison of WIMP-like Freeze-Out vs. UFO}{\label{EMA}}

\begin{figure}[!ht]
\centering
\vskip .2in
\includegraphics[width=3.3in]{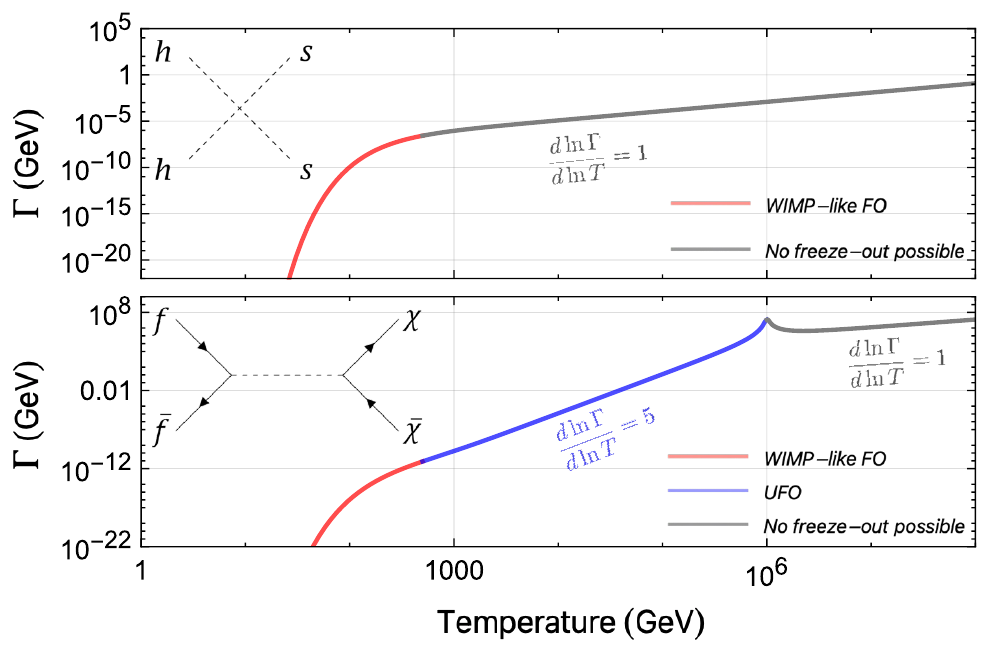}
\caption{\em \small Schematic illustration of the regimes in which WIMP-like FO or UFO can occur for a contact interaction between DM and SM scalars (top) or an interaction via a heavy mediator of mass $M_{s}=10^6$~GeV (bottom), with $m_{\chi}=1$~TeV.}
\label{FreezeOutProperties}
\end{figure}

The temperature-dependence of the thermally averaged cross section (or alternatively, the rate $\Gamma$) in the ultra-relativistic regime determines whether a given particle physics process will be compatible with UFO, according to Eqs.~(\ref{Eq:UFOcondition}) and (\ref{UFOnCondition}). The precise value of $n$ that is required for UFO also depends on the cosmological epoch under consideration, via the temperature dependence of $H$. Cross sections with sufficiently steep temperature dependence will be compatible with UFO, while those with mild temperature dependence will be incompatible. The distinction between these two types of cross sections can be illustrated pictorially by highlighting the distinct 
temperature regimes of $\Gamma(T)$ in which an intersection with $H(T)$ would 
correspond to non-relativistic (WIMP-like) or ultrarelativistic freeze-out, 
which we summarize in Fig.~\ref{FreezeOutProperties} for $m_{\chi}=1$~TeV. The two representative interactions we consider are a contact interaction between DM scalars and the Higgs, as well as fermionic DM production via exchange of a heavy mediator. If $H$ were to intersect $\Gamma$ in the red (blue) regions, WIMP-like FO (UFO) would occur. If $\Gamma < H$ for all $T$ for either interaction, the DM would be FIMP-like. Freeze-out is not possible in the gray regions for standard cosmological eras, specifically for any era with $H(T) \propto T^{\gamma_{2}}$ and $\gamma_{2}>1$. We note that while we have used temperature as the dynamical variable to express the UFO condition in Eqs.~(\ref{Eq:UFOcondition}) and ~(\ref{UFOnCondition}), it is straightforward to express equivalent conditions using either the scale factor, $a$, or time as the dynamical variable. 

\subsection{UFO Parameter Space}{\label{EMParamSpace}}
In this section, we obtain analytic estimates of the UFO parameter space consistent with the observed relic density ($\Omega_{\chi} h^2=0.12$). Specifically, we obtain a range of parameter values for $\Lambda$, $m_{\chi}$, and $\Trh$ such that DM will freeze out ultra-relativistically, become cold before structure formation, and account for the entire observed relic abundance. For specificity, we consider a BSM interaction with a cross-section in the ultra-relativistic regime corresponding to $n=2$ (see Eq.~(\ref{crossSection})) as in the case of a FSF or FVF interaction, and a perturbative reheating period characterized by a Hubble parameter of the form in Eq.~(\ref{HubbleEMD}).

\begin{table*}[ht!]
\centering
\begin{tabular}{l c l}
\hline
\textbf{DM mass} & \textbf{$\Lambda$} & \hspace{20 mm} \textbf{$T_{\rm RH}$} \\
\hline
5 keV            & $7.1\times10^{2}-4.3 \times 10^{13}$ GeV & \hspace{5 mm} $4 \text{ MeV}-9.7\times10^{11}$ GeV \\
100 keV         & $1.5\times10^{3}-1.1 \times 10^{11}$ GeV & \hspace{5 mm} $4 \text{ MeV}-1.2\times10^{8}$ GeV \\
1 MeV         & $2.7\times10^{3}-1.1 \times 10^{9}$ GeV & \hspace{5 mm} $4 \text{ MeV}-1.2\times10^{5}$ GeV \\
100 MeV         & $7.7\times10^{2}-9.6 \times 10^{4}$ GeV & \hspace{5 mm} $6.5 \text{ MeV}-0.10$ GeV \\
1 GeV         & $3.5\times10^{3}-2.8 \times 10^{5}$ GeV & \hspace{5 mm} $41 \text{ MeV}-0.50$ GeV \\
100 GeV         & $6.9\times10^{4}-2.3 \times 10^{6}$ GeV & \hspace{5 mm} $1.6-12$ GeV \\
1 TeV         & $3.1\times10^{5}-6.8 \times 10^{6}$ GeV & \hspace{5 mm} $10-60$ GeV \\
100 TeV         & $6.2\times10^{6}-5.7 \times 10^{7}$ GeV & \hspace{5 mm} $4.1 \times10^{2}-1.5\times 10^{3}$ GeV \\
1 PeV         & $2.7\times10^{7}-1.6 \times 10^{8}$ GeV & \hspace{5 mm} $2.6\times10^{3}-7.2\times10^{3}$ GeV \\
\hline
\end{tabular}
\caption{Parameter space consistent with UFO producing cold DM and the correct relic abundance ($\Omega_{\chi}h^2=0.12$) for several choices of the DM mass, given an $n=2$ interaction (e.g. FSF or FVF) and a reheating period characterized by a Hubble parameter of the form in Eq.~(\ref{HubbleEMD}).}
\label{tab:paramSpace}
\end{table*}

From Eq.~(\ref{TUFO}), we know the ultra-relativistic freeze-out temperature. The upper bound on $T_{\rm FO}$ is determined by $T_{\rm FO} \approx M$; this corresponds to the UFO/FI boundary (see Fig.~\ref{FreezeOutProperties}). The lower bound for UFO during reheating is at $T_{\rm FO} = T_{\rm RH}$ if $m_{\chi}<\Trh$ (boundary between UFO during vs. after reheating) or near $T_{\rm FO} \approx m_{\chi}$ if $m_{\chi}>\Trh$ (UFO/WIMP boundary). The precise boundaries will depend on the width of the massive mediator (for the UFO/FI boundary) as well as lower order terms in $T$ in the thermally averaged cross section (for the UFO/WIMP boundary). The UFO/WIMP boundary in particular is quite model-dependent, so we leave an in-depth analysis for future work. Here, it will suffice to estimate the UFO/WIMP boundary using $T_{\rm FO}\approx m_{\chi}$. Thus, we are interested in the regime $m_{\chi} < T_{\rm FO} < M$ for values of $M$, $\Trh$ and $m_{\chi}$ consistent with $\Omega_{\chi}h^2=0.12$. For simplicity, in this section we take $\Lambda=M$ and it is straightforward to generalize these results.

For $n=2$ and a reheating period characterized by Eq.~(\ref{HubbleEMD}), it is possible to show \cite{HGMO} that the relic abundance for UFO during reheating with $m_{\chi} < \Trh$ is given by 
\begin{equation}
\Omega_\chi h^2\simeq \left(\frac{g_{\chi} \zeta(3)}{\pi^2}\right)^2\frac{2}{\sqrt{3 \alpha}} \frac{\Trh^3 M_{P}}{\Lambda^4}m_{\chi}(5.88\times10^6 \text{ GeV}^{-1})
\,,
\label{Eq:omegak2n2mlesstrh}
\end{equation}

\noindent while the relic abundance for $m_{\chi} > \Trh$ for this interaction type and reheating period is given by
\begin{equation}
\Omega_\chi h^2\simeq \left(\frac{g_{\chi} \zeta(3)}{\pi^2}\right)^2\frac{2}{\sqrt{3 \alpha}} \frac{\Trh^7 M_{P}}{\Lambda^4 m_{\chi}^3} (5.88\times10^6 \text{ GeV}^{-1}),
\label{Eq:omegak2n2mGtrh}
\end{equation}
\noindent where $g_{\chi}$ is the number of internal degrees of freedom of the DM, $\alpha=\frac{\pi^2}{30}g^*$, and $g^*$ is the number of relativistic degrees of freedom in the SM radiation bath. We can then fix $\Omega_{\chi}h^2=0.12$ and use Eqs.~(\ref{TUFO}),(\ref{Eq:omegak2n2mlesstrh}), and (\ref{Eq:omegak2n2mGtrh}) along with the constraint $m_{\chi} < T_{\rm FO} < M$ to determine the allowed parameter space. Recall that we found $\approx 5$ keV to be the lower bound of the DM mass such that UFO DM will be cold before structure formation (for $n=2$) \cite{HGMO}, and recall that the lower limit on $\Trh$ from BBN is 4 MeV.

Using this approach, we obtain a range of $\Lambda$ and $\Trh$ values consistent with UFO producing the correct relic abundance of cold DM, for each choice of $m_{\chi}$. We display the allowed values of $\Lambda$ and $\Trh$ for several choices of $m_{\chi}$ ranging from 5 keV to 1 PeV in Table \ref{tab:paramSpace}. The first striking observation is that there is an enormous range of DM masses, BSM interaction strengths, and reheating temperatures that are compatible with the UFO mechanism of DM production. We also see that the allowed interaction strengths (characterized by $\Lambda$) for UFO are precisely those that lie between typical WIMP and FIMP interaction strengths, consistent with the other results in this paper. Next, we see that light DM ($\approx 5 \text{ keV}-10 \text{ MeV}$) is the most unconstrained for this mechanism, with very large ranges of allowed $\Lambda$ and $\Trh$. However, heavier DM masses are also permitted, but require lower reheating temperatures $\Trh \lesssim 10^4$~GeV and a narrower range of $\Lambda$ values. While the highest $\Trh$ value in Table~\ref{tab:paramSpace} is $\approx 10^{12}$ GeV, and the lowest mass is 5 keV, we show in our companion paper that for a different choice of reheating model (where the inflaton oscillates about the minimum of a quartic rather than quadratic potential), the maximum $\Trh$ and minimum DM mass are $\approx 10^{15}$ GeV and $\approx 200$ eV respectively. Taking all of this into account, we arrive at the range of parameter values we reported in the abstract. 
It is worth observing that although UFO occurs during reheating, this mechanism does not require particularly low reheating temperatures. As a result, it is not difficult for UFO to be compatible with standard mechanisms of baryogenesis.

Lastly, we comment on the prospects for direct detection of UFO DM. In recent years, the null results from WIMP detection experiments based on nuclear recoil have motivated alternative approaches. One of the limitations of nuclear recoil experiments is the ionization threshold, which makes these experiments largely blind to DM with sub-GeV masses. One promising method for probing sub-GeV DM is skipper CCDs, which are ultra-sensitive detectors that rely on electron recoil, such as SENSEI \cite{SENSEI:2025} and DAMIC-M \cite{Aggarwal:2025}. These searches are currently probing DM masses from about several hundred keV to 1 GeV. As can be seen from Table \ref{tab:paramSpace} and from Figs. 7 and 8 in our companion paper \cite{HGMO}, this sub-GeV mass range is precisely the regime which is most optimal for UFO DM. In particular, UFO DM masses from about 5 keV to $\approx 30$~MeV tend to be the least constrained with respect to $T_{\rm RH}$ and $\Lambda$ (depending on the choice of $k$ and $n$), although a much broader range of masses are possible. 

While Table~\ref{tab:paramSpace} provides a window into the allowed parameter space for UFO, we provide a more thorough analysis in a companion paper \cite{HGMO}. The reader should note that Table~\ref{tab:paramSpace} does not exhaust the entire parameter space for UFO during reheating for the indicated DM masses. Instead, in Table ~\ref{tab:paramSpace} we chose a specific interaction type ($n=2$) and reheating model ($H(T)\propto T^{4}$) and $\mathcal{O}(1)$ couplings. For different choices, the allowed ranges for $\Lambda$ and $\Trh$ may expand or contract significantly, testifying to the richness of possibilities enabled by the UFO mechanism.
\end{document}